\documentclass[twocolumn,floatfix,superscriptaddress,english]{revtex4}
\usepackage[T1]{fontenc}
\usepackage[latin9]{inputenc}
\usepackage{color}
\usepackage{soul}
\usepackage{multirow}
\usepackage{graphicx}
\usepackage{calc}
\usepackage{amsmath}
\usepackage{babel}
\usepackage{float}
\usepackage{url} 
\def\code#1{\texttt{#1}}
\usepackage[paperwidth=210mm,paperheight=297mm,centering,hmargin=2cm,vmargin=2.5cm]{geometry}
\usepackage{xargs}                     
\usepackage[pdftex,dvipsnames]{xcolor} 
\usepackage{booktabs}
\graphicspath{{figs/}}
\makeatletter

\usepackage{babel}
\makeatother
\begin{document}
\title{Virtual Pulse Reconstruction Diagnostic for Single-Shot Measurement of Free Electron Laser Radiation Power}
\author{Till Korten }
\affiliation{Helmholtz-Zentrum Dresden-Rossendorf HZDR, 01328 Dresden Germany}
\author{Vladimir Rybnikov} 
\affiliation{Deutsches Elektronen-Synchrotron
DESY, 22607 Hamburg, Germany}

\author{Peter Steinbach}
\affiliation{Helmholtz-Zentrum Dresden-Rossendorf HZDR, 01328 Dresden Germany}
\author{Najmeh Mirian}
\email{najmeh.mirian@hzdr.de}
\affiliation{Helmholtz-Zentrum Dresden-Rossendorf HZDR, 01328 Dresden Germany}
\begin{abstract}

Accurate characterization of radiation pulse profiles is crucial for optimizing beam quality and enhancing experimental outcomes in Free Electron Laser (FEL) research. In this paper, we present a novel approach that employs machine learning techniques for real-time virtual diagnostics of FEL radiation pulses. Our advanced artificial intelligence (AI)-based diagnostic tool utilizes longitudinal phase space data obtained from the X-band transverse deflecting structure to reconstruct the temporal profile of FEL pulses in real time. Unlike traditional single-shot methods, this AI-driven solution provides a non-invasive, highly efficient alternative for pulse characterization.
By leveraging state-of-the-art machine learning models, our method facilitates precise single-shot measurements of FEL pulse power, offering significant advantages for FEL science research. This work outlines the conceptual framework, methodology, and validation results of our virtual diagnostic tool, demonstrating its potential to significantly impact FEL research.
\end{abstract}

\maketitle
\section{Introduction}
In recent years, electron beam accelerators are essential in fields such as medical therapies, materials science, and particle physics research. Their successful operation relies on the precision and stability of the electron beam, which requires advanced diagnostic technologies to monitor and maintain beam quality. Traditional diagnostic methods, while effective, often struggle with the complex, dynamic nature of electron beams. This is where machine learning (ML) emerges as a transformative technology, elevating diagnostic capabilities by providing enhanced precision, adaptability, and efficiency \cite{ratner2020,Kaiser2024, Kaiser2024_2, Fujita2021}. Machine learning algorithms excel at processing and analyzing vast amounts of data from diagnostic sensors and instruments with unprecedented speed and accuracy. These algorithms can identify patterns and anomalies that might indicate issues with the beam or the accelerator components. For instance, ML models can be trained on archived (history) data to recognize normal operating parameters of an electron beam. By continuously monitoring real-time data, these models can quickly detect deviations from the norm, signaling potential issues before they escalate. Techniques such as anomaly detection algorithms and clustering methods allow for the identification of subtle, often imperceptible changes in beam characteristics.\\
The integration of machine learning into diagnostic technology brings numerous advantages: enhanced precision, real-time analysis, adaptability, and data-driven insights. ML algorithms enable highly accurate detection and monitoring of beam characteristics, ensuring precision in diagnostics. The real-time data processing capabilities of ML models allow for the immediate identification of issues and swift corrective actions. Additionally, machine learning systems adapt to new data and evolving conditions, offering a flexible solution for the dynamic nature of electron beams. Moreover, ML-driven diagnostics provide deeper insights into accelerator behavior and performance, facilitating informed decision-making and optimization strategies.\\
As machine learning continues to evolve, its expanding role in the diagnostic technology of electron beam accelerators is gradually transforming the field of accelerator diagnostics and control systems. By improving the precision, reliability, and efficiency of diagnostic processes, ML contributes significantly to the optimal performance and longevity of these complex systems. As these technologies evolve, their impact on diagnostic applications is expected to expand, driving further innovations and enhancements in electron beam accelerator operations \cite{VD2, Vd1-Convery, VD4-Hanuka2021}. \\
Building on these advancements in accelerator technology, Free Electron Lasers (FELs) have revolutionized the generation of highly coherent, powerful electromagnetic radiation at subatomic wavelengths. Innovations in electron accelerators now enable FELs to produce intense, ultrashort pulses, spanning from extreme ultraviolet to hard X-rays at exceptionally high repetition rates. These FELs, powered by high-brightness electron beams, have opened up new possibilities for exploring molecular and atomic dynamics, with profound implications across fields such as physics, chemistry, biology, medical physics, and materials science. The ability to generate coherent, powerful ultrashort pulses in the short-wavelength regime has become crucial for advancing research in these domains.\\
To analyze FEL experimental data, especially for nonlinear interactions, it is essential to have precise knowledge of the FEL pulse peak power and the electric field in the time domain. Conducting pulse-to-pulse measurements of the X-ray pulse structure over time could pave the way for new scientific discoveries. Specifically, having an accurate understanding of the complete electric field of the X-ray pulse, including both its amplitude and phase, is invaluable for many FEL experiments. For instance, more powerful X-ray analogues of optical nonlinear spectroscopies or the resolution-limited standard pump-probe experiments would greatly benefit from precise knowledge of the FEL radiation's electric field.\\
To achieve the precision required in FEL experiments, various diagnostic methods have been developed. One such method is the transverse reconstruction of the electron beam and X-ray algorithm, which aims to reconstruct the temporal power profile P(t) by analyzing the product of the measured current profile and the difference between the energy spread or mean energy profiles of lasing-on and lasing-off shots using a transverse deflecting cavity (TDS)  (see Figure 1 in \cite{SLAC} ). However, this approach is constrained by the finite resolution of the TDS \cite{TDS}, and it is also impossible to simultaneously measure the lasing-on and lasing-off electron phase spaces for a single shot \cite{Florian2020, Hanuka2021}. \\ 
To overcome the single shot measurement challenges, we introduce the Virtual Pulse Reconstruction Diagnostic (VPuRD), a novel tool designed to reconstruct or measure the FEL pulse on a shot-to-shot basis. VPuRD leverages advanced machine learning (ML) techniques to bypass the cumbersome limitations associated with traditional methods of single-shot phase space measurements. Typically, the process involves turning off the lasing for each measurement and capturing the longitudinal phase space across multiple electron beam shots, which is both time-consuming and invasive \cite{SLAC}. The VPuRD tool addresses this by employing ML models to reconstruct the longitudinal phase space of the electron beam using parameters derived from non-invasive diagnostic tools. This innovation enables our virtual diagnostic tool to accurately reconstruct the FEL pulse for each individual electron beam, providing a more efficient and precise measurement process. By eliminating the need for repeated lasing-off measurements, VPuRD not only simplifies the diagnostic procedure but also enhances the accuracy and reliability of FEL pulse characterization.
To demonstrate the applications and advantages of our VPuRD tool, we selected the Free electron LASer in Hamburg (FLASH), a state-of-the-art facility located at Deutsches Elektronen-Synchrotron (DESY) in Germany. In the following section, we will discuss the facility and hardware tools utilized for our measurements. The subsequent sections will be dedicated to the methodology, discussion, and results of our virtual diagnostic tool.
\section{facility, hardware and software}
\subsection*{FLASH}
 FLASH is a free-electron laser machine that generates extremely bright, ultra-short pulses of laser light in the ultraviolet (UV) and soft X-ray regions of the electromagnetic spectrum. This unique capability makes FLASH an invaluable tool for researchers across a wide range of scientific disciplines, including physics, chemistry, biology, and materials science \cite{rossbach2019,flash2020p}. FLASH is a high repetition at 1 MHz in burst mode, i.e
 offers the ability to deliver a burst of 500 electron bunches in each 0.1 second.
FLASH operates by accelerating a train of electron bunches using a superconducting linear accelerator (Linac). These high-energy electrons are then directed through a series of undulator magnets, whose magnetic structures force the electrons to follow a sinusoidal path, causing them to emit synchrotron radiation. Through a process known as self-amplified spontaneous emission (SASE), this radiation is amplified to produce intense, coherent laser pulses. FLASH has two undulator magnet chains, referred to as FLASH1 and FLASH2.\\
\begin{figure*}
    \centering
    \includegraphics[width=0.9\textwidth]{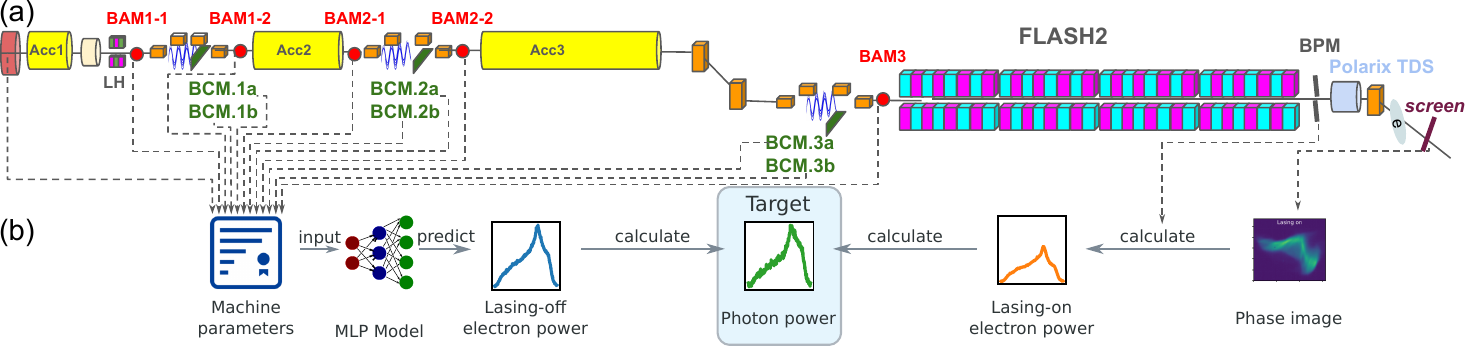}
    \caption{(a) Schematic layout of the FLASH0 (injector and accelerator) and the FLASH2 beamline at DESY.
    Data sources for training the machine learning model are indicated by dashed arrows. (b) Process workflow. During machine operation, we get measured machine parameters and the longitudinal phase space (Phase image) for each electron bunch. We use a multi-layer perceptron machine learning model (MLP Model) to predict the temporal power profile of the electron bunch in the lasing-off condition (Lasing-off e-beam power). From the phase space image, we calculate the temporal power profile of the electron bunch in the lasing-on condition (Lasing-on e-beam power). Thus, we can estimate the temporal power profile for each individual photon pulse (Photon power). }
    \label{fig:FLASH_layout}
\end{figure*}
Figure \ref{fig:FLASH_layout} shows the layout of the common FLASH0 (injector and accelerator) and the FLASH2 FEL beamline. The accelerator section includes two bunch compressor sections, BC1 and BC2, each with two bunch compressor monitor systems (BCMs) and two bunch arrival time monitors (BAMs) at the entrance and exit of the bunch compressors. The FLASH2 line features an additional bunch compressor, allowing for further compression of the electron beam, and includes two BCM systems along with one BAM.\\
The bunch compressor monitor (BCM) is non-invasive diagnostic tool that plays a pivotal role in monitoring the compression of electron bunches to ensure they reach the desired length and characteristics. To achieve precise measurements, the BCM system may integrate various diagnostic tools, including pyroelectric detectors, which are sensitive to infrared radiation. By leveraging the response of pyroelectric materials to temperature fluctuations caused by the interaction of the electron beam with its environment, the BCM can effectively gauge the dynamics of the compressed bunches. The FLASH machine employs both more sensitive and less sensitive pyroelectric detectors within its BCM systems, allowing for enhanced measurement capabilities and improved control over the electron bunch compression process.\\
The Bunch Arrival Monitor (BAM) is also a non-invasive diagnostic tool used to measure the entry and exit times of electron bunches. The longitudinal dispersion in the bunch compression chicanes correlates changes in beam energy upstream of the chicane with arrival time changes downstream of the bunch compressor. In other words, for a given fixed machine setup, the BAM measures the relative energy of the beam during the run.\\
The data acquisition (DAQ) system in our virtual diagnostic setup collects shot-to-shot data from the BAMs and BCMs, as well as the charge and energy of the electron beam at the end of the Linac. We will discuss our DAQ system in detail in the following two sections.\\
One of the most notable features of FLASH is its capability to generate extremely short pulses, on the order of femtoseconds $10^{-15}$ seconds), particularly in the FLASH2 undulator line. These ultra-short pulses allow researchers to study phenomena occurring on extremely fast timescales, such as the dynamics of chemical reactions and electron behavior in materials. Users of the FLASH2 beamline are often interested in the FEL pulse profile. To a certain extent this need is addressed by by a feature of the variable-polarization X-band transversely deflecting structure (POLARIX TDS) downstream of the FLASH2 undulator line \cite{POLARIX}. This device enables pulse reconstruction of the FEL through the transverse reconstruction of the electron beam and X-ray algorithm \cite{Mirian}. We will explore the POLARIX TDS and its role in our virtual diagnostic in detail in a separate subsection.
\subsection*{POLARIX TDS}
The POLARIX TDS is an advanced device that plays a crucial role in the detailed characterization of the electron bunches produced by the FLASH facility. It provides valuable insights into the temporal and spatial properties of these ultra-fast electron beams. 
POLARIX operates at a specific radiofrequency (RF), with almost 12 GHz frequency that is synchronized with the electron bunches passing through it \cite{POLARIX2, POLARIX3, POLARIX4}. The frequency is carefully chosen to match the operational parameters of FLASH2. The cavity generates an electromagnetic field that imparts a transverse momentum to the electrons. Both the strength (RF voltage) and direction (RF phase) of this deflection are meticulously controlled to achieve the desired temporal-to-spatial mapping. This mapping allows for measurements of bunch length and temporal distribution at a resolution of the order of femtoseconds. The device plays a crucial role in optimizing FLASH's performance \cite{POLARIX}.\\
Consequently, the POLARIX TDS is integral to our virtual diagnostic tool. During data collection, we gather information on the POLARIX TDS RF phase, RF voltage, electron beam energy, charge, and the position of the electron both before and after the TDS cavity. Additionally, the longitudinal phase space of the electron beam is captured by the YAG screen located after the bending magnet positioned downstream of the TDS cavity (see Figure \ref{fig:FLASH_layout}). During data acquisition, the time calibration factor was 1.13 fs/mm and energy calibration factor was 21 kev/mm. The time resolution was 15.4 fs in our measurements. 
\subsection*{Data acquisition system}
 The fast Data Acquisition system (DAQ) \cite{DAQ} of FLASH control system has been used by our virtual diagnostic. The DAQ system was developed to
study, monitor and document the machine performance
and parameters and also to collect the results of the
experiment measurements at FLASH. The  DAQ data (machine parameters) is collected in real time with the
unique identification for each shot in the Linac coming from the FLASH timing system.
In this way one can easily correlate any diagnostics DAQ channel data on the bunch-by-bunch level. The collected data can be processed both on and off-line. The first approach is used by the slow feedback middle layer servers.
The second approach makes use of the DAQ data raw files written during the data collection.
The DAQ files are written in a highly optimized custom format. For our virtual diagnostics the raw files were converted to HDF5 files.
\section{Methodology}
Machine learning (ML) methods are increasingly being employed to enhance the accuracy of longitudinal phase space measurements of electron beams in FELs like FLASH2 at DESY. These techniques provide sophisticated tools to analyze complex data and improve the precision of measurements, which are critical for optimizing the performance of FELs \cite{Zhu_2022}.\\
We collected data while the FLASH machine was optimized for FEL radiation for user delivery. The electron beam, with a charge of 200 pC, was accelerated to an energy of 875 MeV, allowing the FLASH2 beamline to generate FEL radiation at 12 nm. In the lasing-off condition, FEL radiation was suppressed by deflecting the electron beam trajectory at the entrance of the undulator line.

For virtual pulse reconstruction, we employed the following workflow (see Figure \ref{fig:FLASH_layout}): For each electron bunch, the DAQ system collected 23 machine parameters (as shown in Table \ref{tab:machine_parameters}) along with a longitudinal phase space image. These machine parameters were used as inputs to predict the shape of the temporal power profile of the electron bunch in the lasing-off condition. The temporal power profile of the electron beam in the lasing-on condition was measured directly from the longitudinal phase space. Consequently, we were able to reconstruct the temporal radiation power profile for each individual photon pulse.\\
Here, we focus on the machine-learning part. Towards this end, we develop a multi-layer perceptron (MLP) machine learning model that is able to predict the temporal power profile of the electron bunch in the lasing-off condition. Towards this end, we collect the machine parameters as well as the longitudinal phase space for 2826 electron bunches in lasing-off condition as training data.
%
%
\subsection*{Data Collection and Preprocessing}
High-resolution data on the electron beam's longitudinal phase space collected by POLARIX TDS at 10 Hz (Figure \ref{fig:jitter}a), preprocessed to remove noise, background and irrelevant information. 
\subsubsection*{Jittering}
\begin{figure}
    \centering
    \includegraphics[width=1\linewidth]{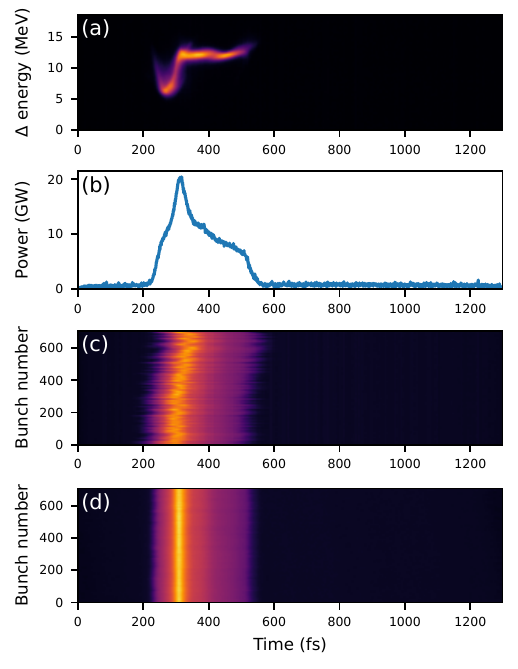}
    \caption{De-jittering. (a) Electron beam phase space image. (b) Temporal power profile created by weighing the phase image by the energy axis and projecting onto the time axis. (c) Temporal power profiles for 700 samples before de-jittering. (d) Temporal power profiles after de-jittering. }
    \label{fig:jitter}
\end{figure}

To compensate for jitter in the electron bunch longitudinal phase space  due to POLARIX radio-frequency jitters (Figure \ref{fig:jitter}), we calculate the electron temporal power profile for each electron bunch (examples shown in Figure \ref{fig:jitter}a and b). The electron bunches show notable temporal jitter (Figure \ref{fig:jitter}c). To compensate for the jitter, the peak power locations are determined from power profiles smoothed by convolving the signal with a Gaussian profile with 10 pixel radius. Then we calculate the offset of each peak to the median peak location and shift the power profiles accordingly; resulting in a nice alignment (Figure \ref{fig:jitter}d). Finally, we crop away the parts of the signal that only contain background in every bunch. The location of the signal is determined by segmenting the signal using Otsu's method \cite{Otsu_1979} and cropping to the bounding box of the segmentation with a padding of 50 pixels. The aligned power profiles are used as labels for the training of the MLP model.
\subsubsection*{Temporal power profile}
The electron temporal power profile is calculated from the charge detected in each slice of the longitudinal phase space image (Figure \ref{fig:jitter}a) multiplied by the corresponding energy difference ($\Delta$energy) in MeV. The resulting energy weighted charge is projected onto the time axis to calculate the electron temporal power profile (Figure \ref{fig:jitter}b).
\subsection*{Model Training and Validation}

We use a simple MLP model with 23 input nodes (see Table \ref{tab:machine_parameters} for details on the input parameters), a single hidden layer with 295 nodes and an output layer with 567 nodes (the width of the electron temporal power profiles in the training data). The 2826 datasets are split into training, validation and test sets with 2261, 283 and 282 samples, respectively.\\
We use an adapted loss function that penalizes regression to the mean:

\begin{equation}
    L = \sum_{i=1}^{D}(x_i-y_i)^2 - \alpha\sum_{i=1}^{D}(x_i-\hat{y})^2
\label{eqn:loss}
\end{equation}
where $\hat{y}$ is the mean of the labels of the entire training dataset, $\alpha$ is a penalty factor, which we set to $0.1$, $x_i$ are the predictions and $y_i$ are the respective labels.\\
The model is trained in Pytorch \cite{Ansel_2024} using the Adam optimizer \cite{Adam_2017}, a dropout of 0.5 on the hidden layer. Training is stopped using the \code{EarlyStopping} callback from Pytorch with a patience of 500. To improve GPU utilization, we use single-batch training, training the model on the entire training dataset Training and validation losses converge well with no indication of overfitting (Figure \ref{fig:ml_results}a).
\begin{table*}
\centering
\caption{Machine parameters used as model input.}
\begin{tabular}{@{}ll@{}}
\toprule
Parameter name & definition\\ \midrule
BCM.1a & measured data from more sensitive pyroelectric detector after BC1\\
norm. BCM.1a & normalized BCM.1a to the bunch charge \\
BCM.1b& measured data from less sensitive pyroelectric detector in BC1 \\
norm. BCM.1b & normalized BCM.1b to the bunch charge\\
BCM.2a & measured data from more sensitive pyroelectric detector after BC2 \\
norm. BCM.2a & normalized BCM.2a to the bunch charge \\
BCM.2b & measured data from less sensitive pyroelectric detector after BC2 \\
norm. BCM.2b & normalized BCM.2b to the bunch charge \\
BCM.3a & measured data from more sensitive pyroelectric detector in FLASH2 after BC3 \\
norm. BCM.3a& normalized BCM.3a to the bunch charge \\
BCM.3b & measured data from less sensitive pyroelectric detector in FLASH2 after BC3 \\
norm. BCM.3b &normalized BCM.3b to the bunch charge  \\
BAM1-1 & bunch arriving time before BC1  \\
BAM1-2 & bunch exciting time after BC1 \\
BAM2-1 & bunch arriving time before BC2 \\
BAM2-2 & bunch exciting time after BC2 \\
BAM3 &  bunch arriving time before BC3 \\
$\Delta t$ (BAM1-2- BAM1-1) & time delay at BC1 \\ 
$\Delta t$ (BAM2-2 - BAM2-1) & time delay at BC2  \\ 
CHARGE in Gun &  electron bunch charge generated at electron gun  \\
CHARGE in FLASH2 &  Electra bunch charge at FLASH2 beamline \\
ENERGY in FLASH2 & electron beam energy\\
BPM x, y & electron beam position before TDS in x and y directions  \\
\bottomrule
     \end{tabular}
    \label{tab:machine_parameters}
\end{table*}

\section{results and discussion}

\subsection{Prediction results}

\begin{figure}[!hbt]
    \centering
    \includegraphics[width=1\linewidth]{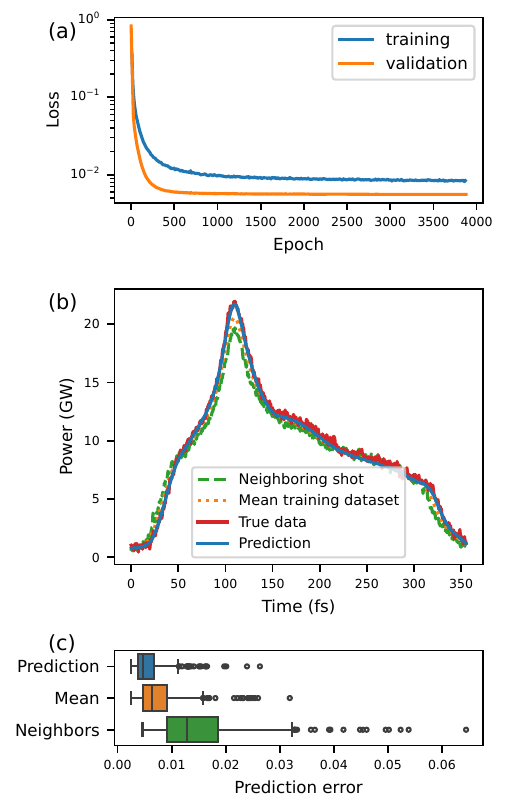} 
    \caption{ MLP model training performance. (a) Training and validation loss. (b)  Predictions for individual shots (blue line) matched the actual measurements (red line) better than measurements from previous shots (dashed green line) and better than the mean of all measurements in the training data (dotted orange line). (c) Boxplots of all mean squared errors in the test dataset. (Prediction) Mean squared error between the predictions and the measurements in the test dataset. (Mean) Mean squared error between the measurements in the test dataset and the mean of all measurements in the training dataset. (Neighbors) Mean squared error between adjacent measurements in the test dataset. 
All three errors are statistically significantly different from each other (Wilcoxon signed-rank test \cite{Wilcoxon_1945} followed by a Bonferroni correction\cite{Bonferroni_1936} for multiple comparisons).}
    \label{fig:ml_results}
\end{figure}

Predicted temporal electron beam power profiles matched the measured profiles very well (Figure \ref{fig:ml_results}b blue line vs. red line). Because of the penalty for predictions that were too close to the mean in the loss function (Equation \ref{eqn:loss}), the model did not simply regress to the mean of the training data set (Figure \ref{fig:ml_results}b orange dotted line). Notably, the neighboring measurement (Figure \ref{fig:ml_results}b green dashed line) fit the measurement even worse than the mean of all shots. These individual observations exemplified in (Figure \ref{fig:ml_results}b) are confirmed by plotting the mean squared errors for all samples in the test dataset (Figure \ref{fig:ml_results}c). The predictions have the lowest mean squared error compared to the measurements in the test dataset ($0.009 \pm 0.0050$ (mean $\pm$ standard deviation, $n=282$)). This is better than the mean squared error between the mean of the entire training dataset and the individual measurements in the test dataset ($0.011 \pm 0.007$). The highest mean squared error was observed between neighboring measurements, can be approximated as
\begin{equation}
    \frac{\sum(x_i - x_{i+1})^2}{n-1}, 
\end{equation}
where $x_i$ are the individual measurements and $n$ is the total number of measurements. In the test dataset ($0.02 \pm 0.014$). All three errors are statistically significantly different to each other ($p<0.01$) as determined by a one-way ANOVA followed by Turkey's HSD test \cite{Tukey_1949}.\\
The observation that neighboring measurements are not good predictors is significant, because neighboring shots have previously been used as labels to train machine learning models to predict longitudinal phase space data \cite{Zhu_2022}.
\subsection{Photon power reconstruction}
\begin{figure}[!ht]
    \centering
    \includegraphics[width=0.95\linewidth]{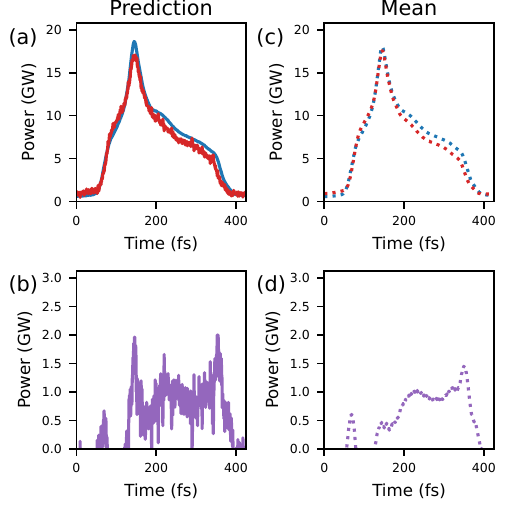}
    \caption{Single shot photon power reconstruction. (a) Electron beam power for one example shot. Electron beam power in the lasing-off condition (blue line) is predicted by the model and electron beam power in the lasing-on (red line) is measured. (b) Reconstructed photon power for the same shot as in (a). (c) Average electron beam power from 2261 shots from the lasing-off condition (blue dotted line) and 574 shots from the lasing-on condition (red dotted line). (d) Average photon pulse power calculated from (c). }
    \label{fig:lasing_on}
\end{figure}

We apply the model trained on 2261 shots in the lasing-off condition to 574 shots acquired in the lasing-on condition (Figure \ref{fig:lasing_on}). The model predicted electron beam powers in the lasing-off condition (blue line in Figure \ref{fig:lasing_on}a). By comparing this with the measured electron beam power in the lasing-on condition (red line in Figure \ref{fig:lasing_on}a), we were able to reconstruct the photon pulse power profile (Figure \ref{fig:lasing_on}b). This shot-to-shot reconstruction is compatible with the state of the art reconstruction method based on averaging the datasets from both the lasing-off and lasing-on conditions (Figure \ref{fig:lasing_on}c and d). \\
As displayed in this figure, the electron beam power in the lasing-on condition in the $80-120$ fs time range is larger than in the lasing-off condition. In pulse reconstruction, this effect results in negative FEL radiation power, which needs to be removed for accurate analysis. This indicates that electrons in this part of the bunch are absorbing FEL radiation, likely due to the slippage effect and phase mismatch. While the electron travel through the undulator line, the high power FEL radiation generated by the peak current (around 170 fs) slips on to the head of electron beam. A portion of this radiation can be absorbed by the electrons, and part of it is observed as FEL radiation at the head of the beam.  
This phenomenon is  evident in the results of both methods. This introduces an uncertainty in the reconstructed pulse's head, though it is explainable and consistent with the slippage mechanism.\\
The temporal spike structure in the SASE (Self-Amplified Spontaneous Emission) spectrum is a well-known characteristic \cite{FEL_chrac}. These spikes result from the stochastic nature of the initial electron beam density modulations that amplify through the undulator. The length of each spike in the time domain (or the spike width) is determined by the coherence time of the FEL radiation. This coherence time, $\tau_c$ , is inversely proportional to the spectral bandwidth $\Delta \omega$ of the SASE radiation $\tau_c\approx 1/\Delta \omega$. For typical X-ray SASE FELs, such as FLASH, this coherence time (and hence the duration of each temporal spike) is in the range of femtoseconds, often between 1 fs to a few tens fs depending on the wavelength and specific machine parameters. At shorter wavelengths, the spikes tend to be narrower, while at longer wavelengths, they can be somewhat broader. In our measurements, FLASH2 was operated at a wavelength of 12 nm, producing temporal spikes of only a few femtoseconds each \cite{FEL_chrac}.
\begin{figure}[!ht]
    \centering
    \includegraphics[width=0.95\linewidth]{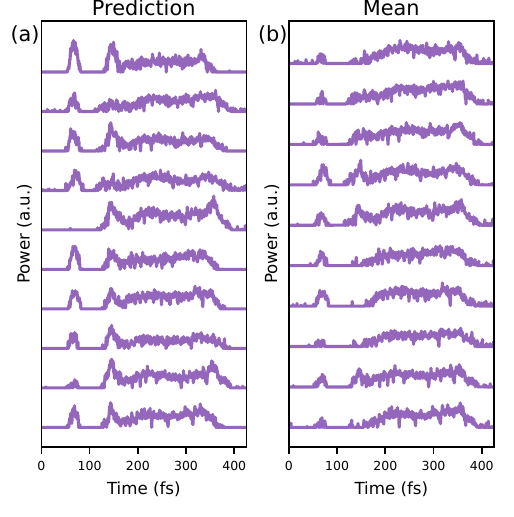}
    \caption{Multiple examples of single shot photon power reconstruction. (a) Photon powers reconstructed from data predicted by the machine learning model (calculated as for Fig. \ref{fig:lasing_on}b). (b) Photon powers reconstructed by subtracting the same measured lasing-on profiles as in (a) from the mean of 2261 shots from the lasing-off condition. }
    \label{fig:multi_signal}
\end{figure}\\
Figure \ref{fig:multi_signal} illustrates ten single-shot FEL pulse reconstructions measured using VPuRD at the FLASH control system. The results are shown for two approaches in plots (a) and (b). In both methods, a single shot of the electron beam phase space in the lasing-on condition directly was used to evaluate the beam power.
In figure \ref{fig:multi_signal}(a)  a machine learning-based method was employed to predict the electron beam power for the lasing-off condition for each FEL shot. In figure \ref{fig:multi_signal}(b) the mean values from 2,261 shots were used to evaluate the electron beam power for the lasing-off condition. As expected, the machine learning method in VPuRD provided a more accurate reconstruction of the FEL pulses.\\
While the VPuRD tool has demonstrated its capability for FEL pulse characterization, it faces a specific challenge that requires consideration. During data collection in the lasing-off condition, the SASE radiation was suppressed by deflecting the electron beam trajectory at the entrance of the undulator line. This approach introduced betatron oscillations and slightly shifted the position of the electron beam on the energy axis.
To address this issue, we compensated for the error using energy measurement data obtained near the screen. In the future, we plan to implement trajectory feedback to correct the beam trajectory after the undulator line. Alternatively, data could be collected by detuning the undulator's resonance condition, which may help mitigate this challenge more effectively.\\ Additionally, it should be noted that slippage presents another challenge for the VPuRD tool. Due to the significant slippage associated with FEL radiation at longer wavelengths, VPuRD tends to provide more accurate results for FEL radiation at shorter wavelengths.
\section{conclusion}
This paper introduces the Virtual Pulse Reconstruction Diagnostic (VPuRD), a novel tool for single-shot reconstruction of FEL radiation power profiles. VPuRD employs a machine-learning-based approach to address key challenges in traditional diagnostic methods, particularly their reliance on extensive lasing-off data. By utilizing longitudinal phase space measurements of the electron beam obtained via the X-band transverse deflection structure (TDS), VPuRD enables accurate reconstruction of temporal power profiles using only lasing-on electron beam data. This makes it a highly efficient and non-invasive diagnostic tool. Its compatibility with high-repetition-rate FEL facilities, such as FLASH, positions it as a significant innovation for real-time diagnostics.\\
In addition to its diagnostic capabilities, VPuRD has the potential to contribute to FEL optimization and feedback systems, offering opportunities to enhance FEL performance and stability.\\
Despite its strengths, VPuRD faces certain challenges, such as the slippage effect, which are particularly pronounced at longer wavelengths and trajectory-induced errors. These effects can introduce uncertainties in certain parts of the reconstructed profiles, limiting the tool's applicability under specific conditions. Addressing these challenges in future developments will further improve VPuRD's accuracy and reliability, broadening its utility for various FEL regimes.\\
In conclusion, VPuRD represents a significant advancement in FEL diagnostics, providing an efficient and reliable method for single-shot pulse characterization. Its implementation has the potential to streamline FEL operations, facilitate experimental success, and enable new opportunities in ultrafast science and related research fields.\\
Looking ahead, we plan to enhance VPuRD with additional capabilities, transforming it into a valuable tool not only for DESY but also for the broader scientific community engaged in accelerator and FEL research.
\section*{Acknowledgments}
The authors like to express the gratitude to the late Siegfried Schreiber, former head of the FLASH facility, for his support and encouragement in the early stages of this project. Particular thanks go to Mathias Vogt, Juliane Roensch-Schulenburg and Stefan Duesterer from DESY for their enlightening discussions, insightful comments, valuable suggestions, and support during FLASH operations.
We also thank members of the FLASH operation team for providing help and conditions to carry out the data collection. We thank the entire Teams of Helmholtz AI Matter and FLASH for invaluable discussions and a great working atmosphere.
 \nocite{*}
\bibliography{main}
\end{document}